\documentclass{ecai} 
\usepackage{microtype}
\usepackage{graphicx}
\usepackage{booktabs}
\PassOptionsToPackage{table}{xcolor}
\usepackage{fancyvrb}

\usepackage{caption} 
\usepackage{tikz}
\usetikzlibrary{shapes.geometric, arrows, patterns}
\tikzstyle{hideblock} = [rectangle, rounded corners, minimum width=1.5cm, minimum height=1cm,text centered, draw=black, fill=green!30]
\tikzstyle{decision} = [ellipse, minimum width=1.5cm, minimum height=1cm, text centered, draw=black]
\tikzstyle{arrow} = [thick,->,>=stealth]

\usepackage{tikz}
\usepackage{amssymb}
\usepackage{subcaption}
\usepackage{tabularx}
\usepackage{algorithm}
\usepackage{algpseudocode} 
\usepackage{multirow}
\usepackage{amsmath}

\newtheorem{definition}{Definition}

\usepackage{amsmath,amsfonts,bm}

\def\eqref#1{equation~\ref{#1}}

\def\1{\bm{1}}

\DeclareMathAlphabet{\mathsfit}{\encodingdefault}{\sfdefault}{m}{sl}
\SetMathAlphabet{\mathsfit}{bold}{\encodingdefault}{\sfdefault}{bx}{n}

\usepackage{url}

\newcommand{\cover}[0]{C}
\newcommand{\secret}[0]{S}

\newcommand{\container}[0]{C'}
\newcommand{\revealsecret}[0]{S'}
\newcommand{\revealsanisecret}[0]{\hat{S}}

\newcommand{\sanitized}[0]{\hat{C}}

\newcommand{\Reveal}[0]{\mathcal{R}}
\newcommand{\Sanitize}[0]{\mathcal{P}}

\newcommand{\Hide}[0]{\mathcal{H}}

\newcommand{\inputim}[0]{x}

\newcommand{\ipSpec}[0]{\text{NCC} \geq 0.95}
\newcommand{\seSpec}[0]{\text{NCC} \leq 0.30}
\newcommand{\fullSpec}[0]{\text{IP } \ipSpec \wedge \text{SE } \seSpec}

\usepackage{xcolor}

\newcommand{\figrefs}[1]{Figure~\ref{#1}} 
\newcommand{\tabref}[1]{Table~\ref{#1}}
\newcommand{\Secrefs}[1]{Section~\ref{#1}} 
\newcommand{\defref}[1]{Definition~\ref{#1}}
\newcommand{\eqrefs}[1]{Equation~\ref{#1}} 

\usepackage[capitalize,noabbrev]{cleveref}
\definecolor{limegreen}{rgb}{0.2, 0.8, 0.2}

\begin{document}


\begin{frontmatter}


\paperid{129} 


\title{Sanitizing Hidden Information with Diffusion Models}


\author[A]{\fnms{Preston K.}~\snm{Robinette}\thanks{Corresponding Author. Email: preston.k.robinette@vanderbilt.edu}}
\author[A]{\fnms{Daniel}~\snm{Moyer}}
\author[A]{\fnms{Taylor T.}~\snm{Johnson}}

\address[A]{Vanderbilt University}

\begin{abstract}
Information hiding is the process of embedding data within another form of data, often to conceal its existence or prevent unauthorized access. This process is commonly used in various forms of secure communications (steganography) that can be used by bad actors to propagate malware, exfiltrate victim data, and discreetly communicate. Recent work has utilized deep neural networks to remove this hidden information in a defense mechanism known as sanitization. Previous deep learning works, however, are unable to scale efficiently beyond the MNIST dataset. In this work, we present a novel sanitization method called DM-SUDS that utilizes a diffusion model framework to sanitize/remove hidden information from image-into-image universal and dependent steganography from CIFAR-10 and ImageNet datasets. We evaluate DM-SUDS against three different baselines using MSE, PSNR, SSIM, and NCC metrics and provide further detailed analysis through an ablation study. DM-SUDS outperforms all three baselines and significantly improves image preservation MSE by 50.44\%, PSNR by 12.69\%, SSIM by 11.49\%, and NCC by 3.26\% compared to previous deep learning approaches. Additionally, we introduce a novel evaluation specification that considers the successful removal of hidden information (safety) as well as the resulting quality of the sanitized image (utility). We further demonstrate the versatility of this method with an application in an audio case study, demonstrating its broad applicability to additional domains.
\end{abstract}

\end{frontmatter}
%
%
\section{Introduction}
Steganography, or the art of hiding information in plain sight, is an area of information hiding that is used for covert communication. In steganography, a secret message is embedded in a seemingly harmless physical or digital medium, termed a cover. A container is the secret and cover together that is then relayed to an intended recipient. Due to the hidden nature of the secret message, communication can occur discreetly between parties without arousing suspicion.

This process can be applied to both physical mediums (e.g., microdots, invisible ink, embedded objects, etc.) as well as digital mediums (e.g., images, videos, audio, text, etc.). The pervasive nature of digital media makes digital steganography a more dangerous communication channel compared to its physical counterpart. Its widespread accessibility and ease of dissemination make digital platforms more susceptible to covert manipulations, potentially enabling malicious actors to spread concealed information or malware at an unprecedented scale and speed. In addition to the widespread effect of malware propagation, steganography via digital mediums can also be used to exfiltrate victim data and communicate with other bad actors. As such, advanced detection and prevention mechanisms are necessary to combat these harmful use cases. While various digital mediums can be used as covers, we focus on the use of images in this work, as these are more accessible and commonly used in the wild.

\begin{figure*}[tbh!]
    \centering
    \includegraphics{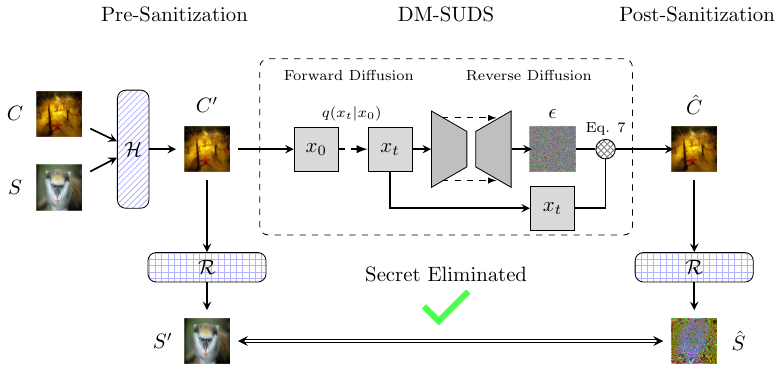}
    \caption{DM-SUDS (center) takes as input a cover or a container image $x_0\in\{\cover, \container\}$ created with any type of steganographic technique. A noisy image is then sampled from this image at timestep $t$. In the reverse diffusion process, a denoising U-Net is used to predict the amount of noise added to the image, which is then used to recover the original image, resulting in a sanitized image $\sanitized$. In the pre-sanitization phase, the secret is recoverable, as demonstrated in the bottom-left of the figure $\revealsecret$. After sanitization with DM-SUDS, however, a secret is not recoverable, as indicated by the bottom-right of the figure $\revealsanisecret$. The secret, therefore, is successfully eliminated.}
    \vspace{1.0em}
    \label{fig:steg_example}
\end{figure*}

Current image steganography defenses utilize a technique known as steganalysis, or the detection of hidden messages in images \cite{johnson1998steganalysis, bachrach2011image}. Steganalysis tools analyze images for known signatures and/or anomalies in pixel values, noise distributions, and other statistical measures to indicate the presence of steganographic content. Recent work has also incorporated machine learning to enhance detection accuracy \cite{zhang2018efficient, ye2017deep, qian2018feature}. These defense strategies, however, rely upon data curated from preexisting steganographic techniques and are referred to as \textit{non-blind}. While adept at detecting known steganography, these defense mechanisms are useless against new forms of steganography in the wild -- especially ones engineered to bypass these existing systems \cite{li2022ensemble,hayes2017generating, tang2019cnn}. 

Another defense mechanism against steganography is sanitization, which eliminates the presence of any potential secret message while maintaining the integrity of the cover media. For instance, a picture of a dog containing an executable malware binary would be sanitized if the malware binary is removed and the original image is minimally changed. In \cite{robinette2023suds}, we demonstrate the success of such an approach, utilizing a variational autoencoder strategy termed SUDS. We show that SUDS is able to protect against least significant bit (LSB), dependent deep hiding (DDH), and universal deep hiding (UDH) steganography, each of which is an unseen method by the sanitizer prior to testing (\textit{blind}). 

While SUDS is able to successfully sanitize secrets, its ability to preserve the quality of the resulting image deteriorates as image complexity increases. This effect is a result of utilizing a variational autoencoder approach. To address the limited preservation capabilities of SUDS while maintaining sanitization performance, we propose a \textbf{d}iffusion \textbf{m}odel approach to \textbf{SUDS}, termed DM-SUDS. While diffusion models are most commonly used for their generative properties, they are trained as denoising mechanisms. By using a diffusion model to ``denoise'' potentially steganographic images, we believe that this approach will provide an improved alternative to SUDS, advancing the state-of-the-art in blind deep learning sanitization techniques for image-into-image steganography. All code to reproduce experiments is available at: \textbf{\url{https://github.com/pkrobinette/dmsuds_steg}}. The primary contributions of this work are the following.%
\begin{enumerate}
    \item \textbf{Implementation of a Novel Sanitization Framework:} We introduce a blind deep learning steganography sanitization method called DM-SUDS that utilizes a diffusion model framework to sanitize image-into-image universal and dependent steganography.
    
    \item \textbf{Demonstration of Sanitizer Capabilities:} We show the benefit of our novel DM-SUDS framework on CIFAR-10 and ImageNet datasets by evaluating it on five different capabilities: comparison to three baselines, flexibility of the timestep parameter (ablation study), ability to directly denoise (ablation study), scalability, and robustness.
    
    \item \textbf{Development of a Novel Sanitization Specification:} We introduce a novel sanitization specification that combines secret elimination (safety) and image preservation (utility). 

    \item \textbf{Application to the Audio Domain}: We present a case study that applies DM-SUDS to a new domain (audio), demonstrating the versatility of this approach applied to text-into-audio steganography. This is the first approach to successfully sanitize information hidden in audio containers using deep learning.
\end{enumerate}%

%
%
\section{Preliminaries}
In this section, we introduce steganography, existing sanitization techniques, and the metrics used for evaluation. While this work is medium agnostic, \textbf{we focus on image-into-image steganography}, where an image secret is embedded into an image cover. An image is represented by a matrix ($c$, $h$, $w$), where $c$ is the number of color channels, $h$ is the height, and $w$ is the width of the image. 

\begin{figure}[b]
    \centering
    \includegraphics[width=\columnwidth]{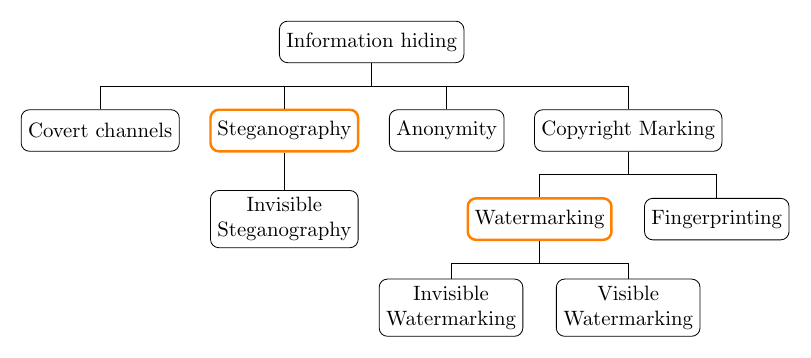}
    \caption{Information Hiding Diagram.}
    \label{fig:info_diagram}
\end{figure}

For the majority of this work, we utilize the term steganography to refer to any covert or imperceptible embedding of information. This includes invisible watermarking. While steganography and watermarking utilize many of the same concepts, some watermarking methods are meant to be visible on the corresponding watermarked media, as shown in \figrefs{fig:info_diagram}. As such, we focus on the removal of steganographic information, which is all invisible, but acknowledge that DM-SUDS is directly applicable to the removal of invisible watermarks as well. 

\subsection{Steganography} 
\textbf{Notation.} As shown by the \textit{Pre-Sanitization} section in \figrefs{fig:steg_example}, steganography typically consists of a \textit{cover}, \textit{secret}, \textit{container}, and a \textit{revealed secret}. A \textit{secret} $\secret$ is hidden within a \textit{cover} $\cover$ using a hide function $\Hide$ to create a \textit{container} $\container$ such that the difference between the cover and the container is minimal, or $\Hide(\cover, \secret) = \container \mid \text{MSE}(\cover, \container) \to 0$. The \textit{revealed secret} $\revealsecret$ can then be obtained from the container using a reveal function $\Reveal$, which is usually the inverse of $\Hide$. The revealed secret should be minimally different from the original secret hidden in the cover, or $\Reveal(\container) = \revealsecret \mid \text{MSE}(\secret, \revealsecret) \to 0$. In regard to sanitization, an image is sanitized with a sanitization function $\Sanitize$ to create a  \textit{sanitized image} $\sanitized$, or $\Sanitize(X) = \sanitized \mid X \in \{\cover, \container\}$. An attempted revealed secret from a sanitized image is denoted as $\revealsanisecret$, or $\Reveal(\sanitized) = \revealsanisecret$. In this work, $\Sanitize \in \{\text{SUDS}, \text{DM-SUDS}, \text{Noise}, \text{DCT-Noise}\}$, where DM-SUDS is our diffusion model approach, SUDS is an existing VAE-based approach, Noise is Gaussian Noise applied directly to the image, and DCT-Noise is Gaussian Noise applied to the Discret Cosine Transform (DCT) of the image.

\textbf{Hiding Methods.}  There are many types of hiding techniques in steganography, including traditional and deep hiding. Traditional hiding involves either 1) modifying the pixels of the image (\textit{spatial domain}) or 2) altering the image using a frequency distribution (\textit{transform domain}) \cite{cheddad2010digital, subhedar2014current, trivedi2016analysis, kishor2016review, holub2012designing}. While information loss is inevitable, the user must strike a balance between maximizing the information hidden in the container and avoiding detection: hiding capacity vs. invisibility. With a higher hiding capacity, a secret is more likely to be detected in a container, but with a lower invisibility, less information is transferable. Traditional methods are marred by this tradeoff, in addition to ensuring robustness against container perturbations. In deep hiding, however, methods are able to incorporate a high capacity of the secret while maintaining invisibility, as these methods are capable of maintaining the pixel distributions of the cover image. Deep hiding techniques utilize deep neural networks as both the hide and reveal functions and fall into two main categories: dependent deep hiding (DDH) \cite{volkhonskiy2020steganographic, yang2019embedding, tang2020automatic, zhu2018hidden, baluja2017hiding, yang2023pris, lu2021large, xu2022robust, jing2021hinet, zhou2019security, zhang2019invisible} and universal deep hiding (UDH) \cite{zhang2020udh}. In DDH, the resulting container is cover dependent, and in UDH, the secret can be combined with any cover.  In this work, we utilize a method from each of these categories 1) traditional = least significant bit (LSB) method \cite{kurak_CSAC}, 2) dependent deep = DDH, and 3) universal deep = UDH. The implementations for DDH and UDH are CNN-based implementations from the same code base\footnote{\textbf{DDH/UDH: }\url{https://github.com/ChaoningZhang/Universal-Deep-Hiding}}, and the LSB implementation is from the SUDS codebase\footnote{\textbf{SUDS  Code:} \url{https://github.com/pkrobinette/suds-ecai-2023}}. This hiding methods were chosen to resemble the previous work from \cite{robinette2023suds}, which is currently the only blind, deep learning method for image-into-image sanitization.


\subsection{Sanitization}
\label{sec:sanitization}
\textbf{Traditional Sanitization.} Sanitization, or active steganalysis, is the process of removing hidden information from potentially steganographic images while preserving the image quality of the cover. Traditional sanitization approaches attempt to remove hidden information by modifying the pixels of the image or the frequency distribution of the image, much like traditional steganographic approaches. Examples include flipping the bits of N LSB planes \cite{paul2010image}, adding Gaussian noise to the image or transformation of the image, and applying image filters \cite{geetha2021steganogram}. While traditional sanitization techniques are effective on traditional hiding algorithms, they largely degrade the image quality of the restored image and are ineffective on deep hiding techniques.

\textbf{Deep Learning Sanitization.}
Deep learning sanitization attempts to improve upon the limited capabilities of traditional sanitization, especially in regard to deep learning steganography. While most work is related to \textbf{text-into-image} steganography and watermarking, recent research has explored \textbf{image-into-image} approaches. This is significant as deep learning methods, which primarily focus on image-into-image steganography, can be adversarially trained to circumvent steganalysis techniques, including Xu-Net \cite{xu2016structural} and Ye-Net \cite{ye2017deep}. In \cite{li2021concealed}, the authors present a GAN-based approach, where a U-Net generator is trained to attack watermarked images. This approach, however, is considered non-blind as it relies on known hiding techniques during the training process. Non-blind techniques are unable to defend against novel hiding strategies, as they are implemented using past, known processes. A more effective approach to sanitization is a blind methodology, where the training process does not incorporate known hiding techniques. In \cite{jung2021pixelsteganalysis}, the authors present a blind approach  called PixelSteganalysis, which uses an `analyzer' to predict pixel and edge distributions of a container, and an `eraser' to sterilize the images by adjusting suspicious pixels indicated by the analyzer. Their approach is inefficient (e.g., a single (3,32,32) CIFAR-10 image takes an average of 3 min. to sanitize) and also requires known information about `high-frequency areas' to train the edge distribution model. In \cite{robinette2023suds}, we introduce another blind sanitization approach, which utilizes a VAE-based approach called SUDS and demonstrate blind sanitization capabilities against unseen traditional, dependent deep, and universal deep hiding methods on MNIST and CIFAR-10 datasets. While this previous approach is able to eliminate secret information from both datasets, the image preservation of the sanitized image starts to deteriorate with CIFAR-10.

\textbf{Diffusion Model Applications.}
In this work, we utilize a diffusion model approach to sanitization. Recent work has applied diffusion models to remove adversarial noise in the spatial domain \cite{nie2022diffusion} and to attack a single \textit{text-into-image} watermarking technique in \cite{li2023diffwa}. 

\subsection{Sanitization Metrics}
\label{sec:image_metrics}
We evaluate sanitization and reconstruction using four image quality metrics: mean squared error (MSE), peak signal-to-noise ratio (PSNR), structural similarity index measure (SSIM), and normalized correlation coefficient (NCC). MSE and PSNR quantify the absolute error between the reference and altered images. SSIM assesses perceptual quality compared to a reference, while NCC measures correlation between images, reflecting their spatial alignment. NCC ranges from -1 to 1 and is crucial in watermarking to verify watermark extraction \cite{vaidya2023fingerprint, quan2020watermarking, zhang2020model}, with 1 indicating identical images. 

Unlike previous methods focusing solely on secret elimination, our approach adds image preservation to assess sanitization quality. This ensures sensitive information is removed without compromising image utility. Drawing from the watermarking domain, we use NCC to evaluate both the preservation and elimination aspects: for image preservation (IP), an $\ipSpec$ indicates successful preservation using the cover image. For secret elimination (SE), an $\seSpec$ signifies successful sanitization, reflecting typical standards for data removal.

\begin{definition}[Sanitization]\label{def:sanitization}
Combining IP and SE, an image is considered successfully sanitized in this work if: $\fullSpec$.
\end{definition}

\section{DM-SUDS Sanitization}
\label{sec:diff_model}
Our goal is to improve the image preservation of sanitized images using a diffusion model approach to sanitization. A diffusion model is a generative model that consists of two main features: a forward diffusion process and a reverse diffusion process. In the forward diffusion process, an input image $x_0$ from a given data distribution $x_0 \sim q(x_0)$ is perturbed at timestep $t$ with Gaussian noise with a variance $\beta_t \in (0, 1)$ to produce a sequence of latent images $\inputim_0, \inputim_1, ..., \inputim_T$. This forward noising process is defined by equations \ref{eq:f_diff1} and \ref{eq:f_diff2} and is commonly reparameterized by \eqrefs{eq:f_diff3}, which allows for direct sampling of noised latents at arbitrary steps. Here, $\alpha_t := 1 - \beta_t$, $\bar{\alpha}_t := \prod_{s=0}^{t} \alpha_s$, and $1-\bar{\alpha}_t$ represents the variance of the noise for an arbitrary timestep.%
\begin{align}
q(x_1, \dots, x_T | x_0) &:= \prod_{t=1}^{T} q(x_t|x_{t-1}) \label{eq:f_diff1} \\
q(x_t|x_{t-1}) &:= \mathcal{N} \left( x_t; \sqrt{1 - \beta_t} x_{t-1}, \beta_t \mathbf{I} \right) \label{eq:f_diff2} \\
q(x_t | x_0) &= \mathcal{N} \left( x_t; \sqrt{\bar{\alpha}_t} x_0, (1 - \bar{\alpha}_t) \mathbf{I} \right) \label{eq:f_diff3}
\end{align}

\noindent The model then learns to reverse this diffusion process by refining the noised sample until it resembles a sample from the target distribution. The posterior $q(x_{t-1}|x_t, x_0)$ can be calculated using Bayes theorem in terms of $\tilde{\beta}_t$ and $\tilde{\mu}_t(x_t, x_0)$, which are defined in the equations below:%
\begin{align}
\tilde{\beta}_t &:= \frac{1 - \bar{\alpha}_{t-1}}{1 - \bar{\alpha}_t} \beta_t \label{eq:r_diff1} \\
\tilde{\mu}_t(x_t, x_0) &:= \frac{\sqrt{\bar{\alpha}_{t-1} \beta_t}}{1 - \bar{\alpha}_t} x_0 + \frac{\sqrt{\alpha_t(1 - \bar{\alpha}_{t-1})}}{1 - \bar{\alpha}_t} x_t \label{eq:r_diff2} \\
q(x_{t-1}|x_t, x_0) &= N \left( x_{t-1}; \tilde{\mu}(x_t, x_0), \tilde{\beta}_t \mathbf{I} \right) \label{eq:r_diff3}
\end{align}

\noindent To represent $\mu_{\theta}(x_t, t)$ for the reverse diffusion process, we use a U-Net model to predict the noise $\epsilon$ added to the input image. The original image can then be predicted using \eqrefs{eq:r_diff4}.%
\begin{equation}
\label{eq:r_diff4}
x_0 = \frac{1}{\sqrt{\alpha_t}} \left( x_t - \frac{\beta_t}{\sqrt{1 - \bar{\alpha}_t}}\epsilon \right)
\end{equation}


While diffusion models are most commonly used for generative purposes, we instead make use of the denoising capabilities developed during training. To use the diffusion model as a sanitizer, we can apply $t$ timesteps of noise to a potential container using \eqrefs{eq:f_diff3} with a cosine beta scheduler from \cite{ho2020denoising}, predict the noise using a neural network, and then refine the image using \eqrefs{eq:r_diff4}, effectively preserving the image quality while maintaining sanitization performance. As this is a blind approach to sanitization, prior knowledge of steganography techniques is not required for the training process. To highlight this feature of the framework and to emphasize the accessibility of this approach, we make use of a state-of-the-art, publicly available pretrained diffusion model from \cite{nichol2021improved}\footnote{Diffusion Model: \url{https://github.com/openai/improved-diffusion}}. For the remainder of this paper, the diffusion model sanitization process is referred to as DM-SUDS. 

\begin{table*}[tb!]
    \caption{Image Preservation (IP) and Secret Elimination (SE) performance metrics on LSB, DDH, and UDH containers made from 1000 CIFAR-10 image pairs. If $\fullSpec$ (\defref{def:sanitization}), this results in a \textcolor{blue}{Successful} sanitization (highlighted in green).}
    \vspace{0.75em}
    \label{tab:combined_comparison}
    \centering
    \small
    \begin{tabularx}{\textwidth}{>{\centering\arraybackslash}p{0.50cm}| >{\centering\arraybackslash}p{1.50cm} 
    p{2.00cm} >
    {\centering\arraybackslash}p{0.75cm} >{\centering\arraybackslash}p{2.25cm} >{\centering\arraybackslash}X >{\centering\arraybackslash}X >{\centering\arraybackslash}X >{\centering\arraybackslash}p{2.20cm}}
        \toprule
        \multirow{2}{*}{\textbf{RQ}} & \multirow{2}{*}{\textbf{Hide}} & \multirow{2}{*}{\textbf{Method}} & \textbf{Time}& \textbf{MSE} & \textbf{PSNR} & \textbf{SSIM} & \textbf{NCC} & \textbf{Sanitization} \\
            & &  & (ms) & (IP / SE)    & (IP / SE)     & (IP / SE)     & (IP / SE)  & (IP / SE)  \\
\midrule
\multirow{5}{*}{RQ1}& \multirow{5}{*}{LSB}  & None     & $4.0$ & 37.47 / 79.15   & 32.50 / 29.17 & 0.97 / 0.97 & 0.99 / 1.00 & - \\
    & & Gaussian Noise & $0.0$ & 430.17 / 9610.91 & 21.80 / 8.40  & 0.74 / 0.01 & 0.92 / 0.02 & \textcolor{red}{Fail} / \textcolor{blue}{Success} \\
    & & DCT Noise  & $0.1$   & 254.81 / 9419.64 & 24.19 / 8.49  & 0.82 / 0.01 & 0.95 / 0.02 & \cellcolor{limegreen}\textcolor{blue}{Success} / \textcolor{blue}{Success} \\
    & & SUDS    & $0.6$      & 359.81 / 9873.25 & 23.04 / 8.29  & 0.78 / 0.01 & 0.93 / -0.00 & \textcolor{red}{Fail} / \textcolor{blue}{Success} \\
    & & DM-SUDS* & $101.4$  & 163.18 / 9200.26 & 26.21 / 8.61  & 0.88 / 0.01 & 0.97 / 0.01 & \cellcolor{limegreen}\textcolor{blue}{Success} / \textcolor{blue}{Success}\\
\midrule
\multirow{5}{*}{RQ1} & \multirow{5}{*}{DDH} & None   & $15.8$ & 114.22 / 93.43   & 28.59 / 29.17 & 0.95 / 0.95 & 0.98 / 0.98 & - \\
    & & Gaussian Noise  & $0.0$ & 516.46 / 3119.53 & 21.07 / 13.29 & 0.71 / 0.30 & 0.90 / 0.50 & \textcolor{red}{Fail} / \textcolor{red}{Fail} \\
    & & DCT Noise  & $0.1$   & 336.37 / 1363.44 & 23.07 / 17.33 & 0.80 / 0.63 & 0.93 / 0.83 & \textcolor{red}{Fail} / \textcolor{red}{Fail}\\
    & & SUDS  & $0.5$       & 391.79 / 5419.22 & 22.71 / 11.16 & 0.78 / 0.08 & 0.92 / 0.24 & \textcolor{red}{Fail} / \textcolor{blue}{Success} \\
    & & DM-SUDS* & $101.8$  & 231.27 / 5082.42 & 24.92 / 11.45 & 0.86 / 0.13 & 0.95 / 0.25 & \cellcolor{limegreen}\textcolor{blue}{Success} / \textcolor{blue}{Success}\\
\midrule
\multirow{5}{*}{RQ1} & \multirow{5}{*}{UDH} & None      & $15.3$ & 23.36 / 155.69   & 34.69 / 26.81 & 0.97 / 0.92 & 0.99 / 0.97 & - \\
    & & Gaussian Noise & $0.0$ & 415.46 / 4793.60 & 21.96 / 11.37 & 0.74 / 0.22 & 0.92 / 0.39 & \textcolor{red}{Fail} / \textcolor{red}{Fail} \\
    & & DCT Noise   & $0.1$   & 243.50 / 905.63  & 24.38 / 18.89 & 0.82 / 0.61 & 0.95 / 0.82 & \textcolor{blue}{Success} / \textcolor{red}{Fail} \\
    & & SUDS  & $0.6$      & 343.71 / 14379.81 & 23.29 / 7.09 & 0.79 / 0.03 & 0.94 / 0.04 & \textcolor{red}{Fail} / \textcolor{blue}{Success}\\
    & & DM-SUDS*  & $101.1$   & 148.40 / 14667.25 & 26.66 / 6.99 & 0.88 / 0.03 & 0.97 / 0.04 & \cellcolor{limegreen}\textcolor{blue}{Success} / \textcolor{blue}{Success}\\
    \midrule
    \multirow{2}{*}{RQ4} & DDH (ImageNet) & \multirow{2}{*}{DM-SUDS*} & \multirow{2}{*}{2191.4} & \multirow{2}{*}{106.42 / 7607.07} & \multirow{2}{*}{28.36 / 9.63} & \multirow{2}{*}{0.84 / 0.07} & \multirow{2}{*}{0.98 / 0.04} & \multirow{2}{*}{\cellcolor{limegreen}\textcolor{blue}{Success} / \textcolor{blue}{Success}} \\
    \midrule
    \multirow{2}{*}{RQ4} & PRIS (ImageNet) & \multirow{2}{*}{DM-SUDS*} &  \multirow{2}{*}{1625.2} & \multirow{2}{*}{89.00 / 4855.80} & \multirow{2}{*}{28.91 / 11.78} & \multirow{2}{*}{0.75 / 0.24} & \multirow{2}{*}{0.96 / 0.30} & \multirow{2}{*}{\cellcolor{limegreen}\textcolor{blue}{Success} / \textcolor{blue}{Success}} \\
    \bottomrule
    \end{tabularx}
\end{table*}

%
%
\section{Research Questions and Metrics}
\label{sec:research_questions}
To evaluate the sanitization performance of DM-SUDS, we seek to answer the following four research questions.

\begin{enumerate}
    \item \textbf{RQ1: Ability to sanitize hidden information $\to$} \textit{Is DM-SUDS able to successfully sanitize hidden information? How does this compare to previous sanitization methods?} To evaluate the performance of DM-SUDS, we compare DM-SUDS against three different baselines: 1) SUDS, 2) Gaussian Noise, and 3) Discrete Cosine Transform (DCT) Noise. We test each of these methods on LSB, DDH, and UDH containers created with the same 1000 image pairs from the CIFAR-10 dataset of size $\mathcal{R}^{3\times32\times32}$. CIFAR-10 is a widely used dataset for information hiding \cite{yang2023general, wang2022stegnet, pan2022house, liu2022deep}, and it is also where the reconstruction capabilities of SUDS start to break down. For the DM-SUDS model discussed in \Secrefs{sec:diff_model}, we use a timestep variable $t=500$ in the forward and reverse diffusion processes, which is half of the number of timesteps used during model training. We utilize the sanitization metrics discussed in \Secrefs{sec:image_metrics} (MSE, PSNR, SSIM, NCC) to evaluate image preservation and secret elimination. To provide a baseline comparison to the effect of each steganographic technique on image quality, a \textbf{None} column is also calculated, which uses the metrics mentioned above to compare \textit{covers} $\cover$ with \textit{containers} $\container$ (\textbf{IP}) as well as \textit{secrets} $\secret$ to \textit{revealed secrets \textbf{pre} sanitization} $\revealsecret$ (\textbf{SE}). \textbf{RQ1} is positive if DM-SUDS successfully sanitizes according to \defref{def:sanitization} while outperforming the baselines in image preservation metrics.

    \item \textbf{RQ2: Effect of timestep on sanitization $\to$} \textit{Does the number of diffusion timesteps affect sanitization performance?} In the previous experiment, we utilize a timestep $t = 500$ in the diffusion process. This means that the noisy version of the input image is sampled from the cosine noise scheduler at $t=500$. In the reverse diffusion process, this number is then encoded, and the neural network estimates the added noise to the image from this timestep. To determine if the number of timesteps affects sanitization performance, we evaluate DM-SUDS on LSB, DDH, and UDH containers using timestep values from $t=25$ to $t=1000$ in 25-step intervals, e.g., $t=[25, 50, 75, ..., 1000]$. We utilize the same container creation process and sanitization evaluation described in RQ1. \textbf{RQ2} is positive if 50\% of the timesteps are successfully sanitized (\defref{def:sanitization}).  

    \item \textbf{RQ3: Ability to directly sanitize $\to$} \textit{Is added noise necessary for the sanitization process?} In RQ1 and RQ2, we sample a noisy image of the container at timestep $t$, and then denoise this image to a sanitized version of the input. Another potential way to sanitize via DM-SUDS is to treat the potentially steganographic image as the direct input to the denoiser, skipping the forward diffusion process. With this approach, however, we still have to provide an estimate $t$ as input to the reverse diffusion neural network. We evaluate the direct sanitization capabilities of DM-SUDS using the evaluation process described in RQ2. \textbf{RQ3} is positive if the result of no noise added resembles that of RQ2 (with noise).

    \item \textbf{RQ4: Scalability and robustness $\to$} \textit{Does DM-SUDS extend to JPEG-resistant steganography and the ImageNet dataset?} As the reconstruction ability of SUDS decreases with an increase in image complexity \cite{robinette2023suds}, we evaluate DM-SUDS on more complex images using the ImageNet Dataset with images of size $\mathcal{R}^{3\times128\times128}$. While DDH is the most robust of the experimental methods, it is mainly used for lossless image types, such as PNG or BMP. Recent work in dependent deep hiding has shown the promise of JPEG (lossy) resistant image steganography as well. As such, we evaluate DM-SUDS on PRIS \cite{yang2023pris}, an invertible network image steganography approach that is robust under JPEG compression. This method was chosen as the authors demonstrate its improved performance against ISN \cite{lu2021large}, RIIS \cite{xu2022robust}, HiNET \cite{jing2021hinet}, and Baluja's dependent deep learning approach \cite{baluja2019hiding}, other competing robust image-into-image steganography techniques. As such, we test DM-SUDS on 500 DDH and PRIS containers created using ImageNet images and collect MSE, PSNR, SSIM, and NCC metrics for secret elimination and image preservation. \textbf{RQ4} is positive if DDH and PRIS containers are successfully sanitized according to \defref{def:sanitization}.

\end{enumerate}

%
%

\section{Experimental Results}
\label{sec:results}
\subsection{Sanitization Performance of DM-SUDS}
\label{sec:results_rq1}
We evaluate DM-SUDS sanitization performance against three different baselines on 1000 containers made from the CIFAR-10 dataset. From the results shown in \tabref{tab:combined_comparison}, DM-SUDS is able to successfully sanitize LSB, DDH, and UDH containers as indicated by the image preservation (IP) and secret elimination (SE) values for each metric. The low MSE, high PSNR, high SSIM, and high NCC for DM-SUDS IP demonstrate that the quality of the sanitized image is preserved, and the high MSE, low PSNR, low SSIM, and low NCC values for DM-SUDS SE validate that the secret has been successfully removed. While previous works only consider SE for successful sanitization, the \textit{Sanitization} evaluation in the last column considers both IS and SE ($\fullSpec$). From these results, DM-SUDS is the only method to successfully sanitize all three unseen hiding methods, outperforming each of the baselines in IP. \textbf{RQ1 is, therefore, positive. }

\begin{figure*}[tph!]
    \centering
    \begin{subfigure}[b]{\columnwidth}
        \centering
        \includegraphics[width=0.75\linewidth]{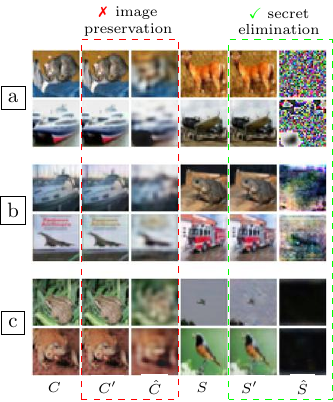}
        \caption{SUDS}
        \label{fig:suds_results}
    \end{subfigure}
    \begin{subfigure}[b]{\columnwidth}
    \centering
        \includegraphics[width=0.75\linewidth]{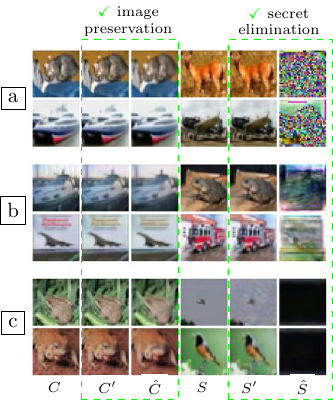}
        \caption{DM-SUDS}
        \label{fig:diffusion_results}
    \end{subfigure}
    \vspace{1.0em}
    \caption{A comparison between DM-SUDS and SUDS sanitization for a) LSB, b) DDH, and c) UDH steganography. Sanitization performance is determined from image preservation as well as secret elimination. While the secrets are eliminated with each method, DM-SUDS is the only method able to preserve the image. }
    \vspace{1.0em}
    \label{fig:results_comparison}
\end{figure*}

The results for SUDS highlight the importance of considering image preservation in addition to secret elimination for sanitization.  According to SE, SUDS effectively removes the secret (low SE NCC), but it also significantly degrades the quality of the resulting image (low IP NCC). This adverse effect is further demonstrated by \figrefs{fig:results_comparison}. In these examples, a cover $\cover$ is combined with a secret $\secret$ to create a container $\container$ using a) LSB, b) DDH, and c) UDH. This container is then sanitized by either SUDS (\figrefs{fig:suds_results}) or DM-SUDS (\figrefs{fig:diffusion_results}) to create a sanitized container $\sanitized$. The sanitized secret $\revealsanisecret$ shows the attempted revealed secret after sanitization. While SUDS and DM-SUDS are both able to eliminate the secret as shown by the last two columns, the image quality of the sanitized container for SUDS is low. Both specifications (IP and SE) are, therefore, necessary to holistically evaluate sanitization.
%
%

\subsection{Effect of Timestep on Sanitization}
In this section, we evaluate the impact of timestep size on the sanitization process. The top two rows of \figrefs{fig:rq2-rq3} show the sanitized imaged $\sanitized$ and an attempted recovered secret $\revealsanisecret$ at various timesteps.  In regard to the sanitized image $\sanitized$, with too many timesteps, the image quality of the reconstructed image starts to deteriorate, as shown by $t=1000$. The blurry nature of the resulting processed image indicates that the more noise that is added to a container, the more difficult it is to predict the amount of added noise, affecting the resulting refined image $\sanitized$. With too few timesteps, however, the secret persists after sanitization. At $t=25$, for instance, the boat of the original secret is still visible in $\revealsanisecret$. These results are further validated by \figrefs{fig:timesteps}, which shows IP NCC and SE NCC for DDH, UDH, and LSB sanitization. The points not in the red bar signify successful sanitization according to \defref{def:sanitization}. As the timestep increases, secret elimination performance improves and image preservation performance deteriorates. These results also offer an interesting insight into the robustness of each hiding method. The order of robustness as determined by the persistence of the secret with increased timesteps is DDH, UDH, and LSB. DDH, therefore, is a more robust hiding method than the other two. While the number of selected timesteps impacts performance near the edges, using a better diffusion model eliminates the need to consider the timestep for sanitization as image preservation is the limiting factor. As less than 50\% of the timesteps result in successful sanitization in this experiment, \textbf{RQ2 is negative}.

%
%
\subsection{Ability to Directly Denoise}
\begin{figure*}[thp!]
    \centering
    \includegraphics[width=\linewidth]{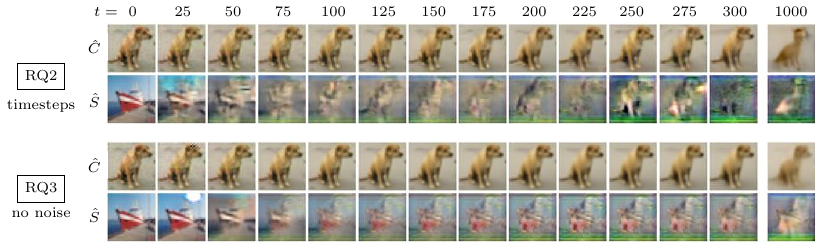}
    \caption{Example of sanitized images $\sanitized$ and attempted revealed secrets after sanitization $\revealsanisecret$ on DDH containers across various timesteps. }
    \label{fig:rq2-rq3}
    \vspace{1.0em}
\end{figure*}

\begin{figure*}[htbp]
    \centering
    \begin{subfigure}{\columnwidth}
        \centering
            \includegraphics[width=0.80\textwidth]{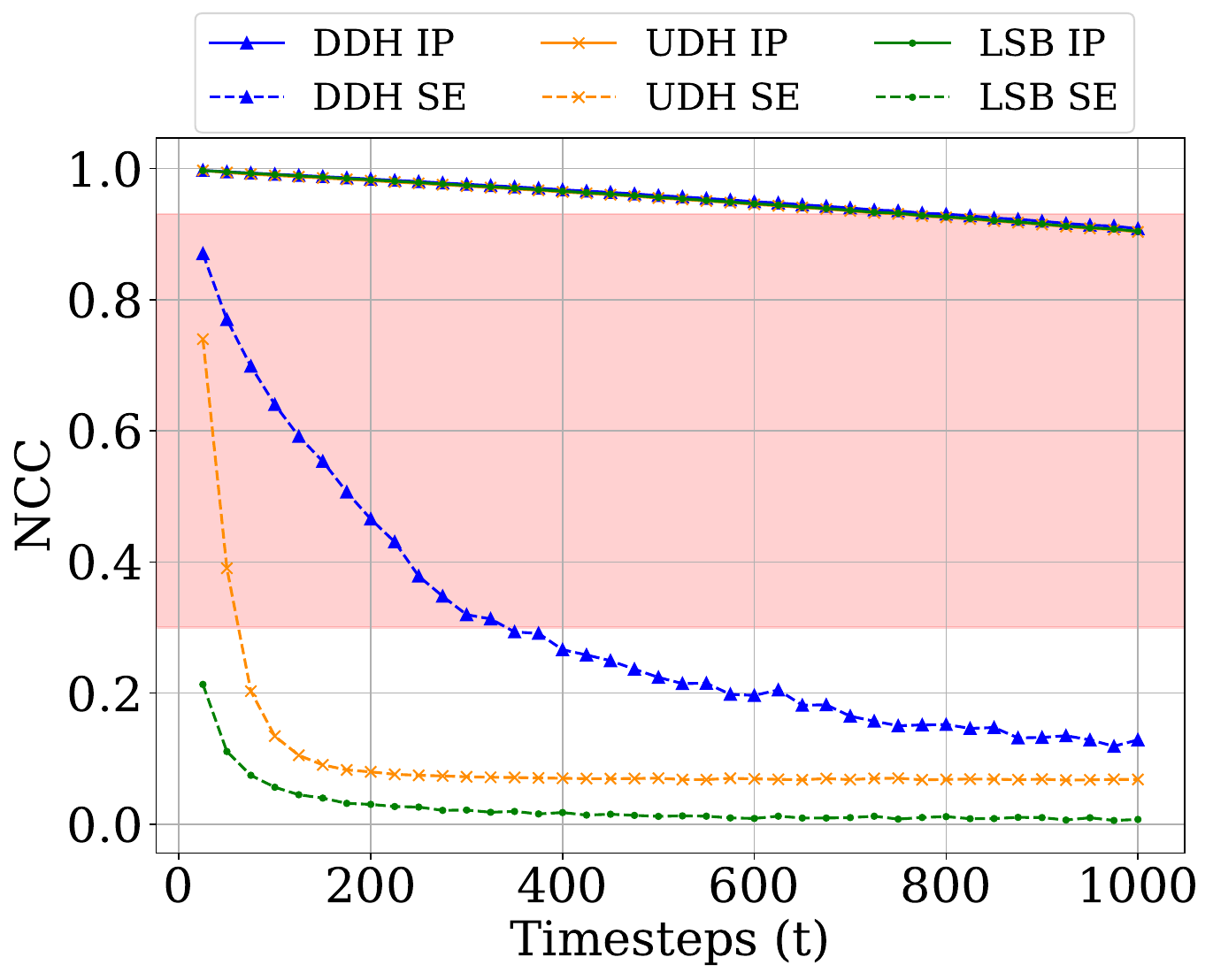}
            \caption{RQ2: Effect of Timesteps}
            \label{fig:timesteps}
    \end{subfigure}
    \begin{subfigure}{\columnwidth}
        \centering
            \includegraphics[width=0.80\textwidth]{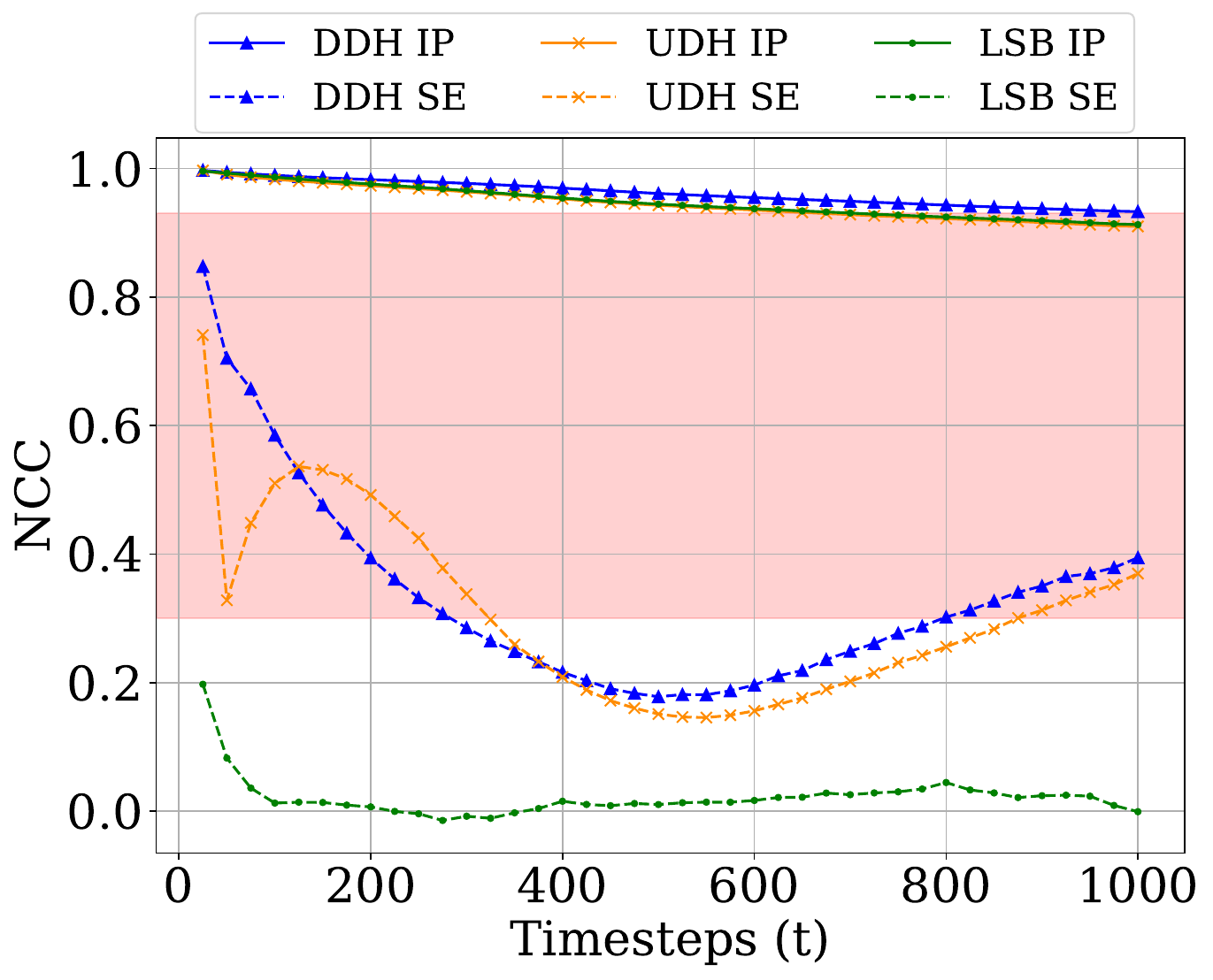}
            \caption{RQ3: Effect of No Added Noise}
            \label{fig:no_noise}
    \end{subfigure}
    \vspace{1.5em}
    \caption{Image preservation (IP) and secret elimination (SE) NCC metrics from sanitizing DDH, UDH, and LSB containers on various timesteps using added noise (a) and no added noise (b). }
    \label{fig:compare_sanitization}
    \vspace{1.0em}
\end{figure*}

The plot in \figrefs{fig:no_noise} shows the IP and SE NCC metrics from sanitizing with no noise added for each hiding method across various timesteps. We see the best sanitization performance at $t=500$. Here, the SE is lowest and the IP is above the indicated threshold. Whereas the SE from sanitizing \textbf{with} added noise continues to decrease with each timestep, the SE with no added noise starts to increase at $t=500$, as shown by \figrefs{fig:compare_sanitization}. So much so, that the secret is not eliminated at higher timesteps. This is further demonstrated by \figrefs{fig:rq2-rq3}, which shows the image results of sanitization $\sanitized$ and an attempted revealed secret after sanitization $\revealsanisecret$ for various timesteps. At $t=1000$, the secret is still discernible from $\revealsanisecret$. These results indicate that the true power of the diffusion model sanitization approach lies in the added Gaussian Noise in the forward diffusion process. As these trends do not resemble those of RQ2, \textbf{RQ3 is negative.}

\begin{figure*}[htb!]
    \centering
    \begin{subfigure}{0.26\textwidth}
        \centering
        \includegraphics[width=\linewidth]{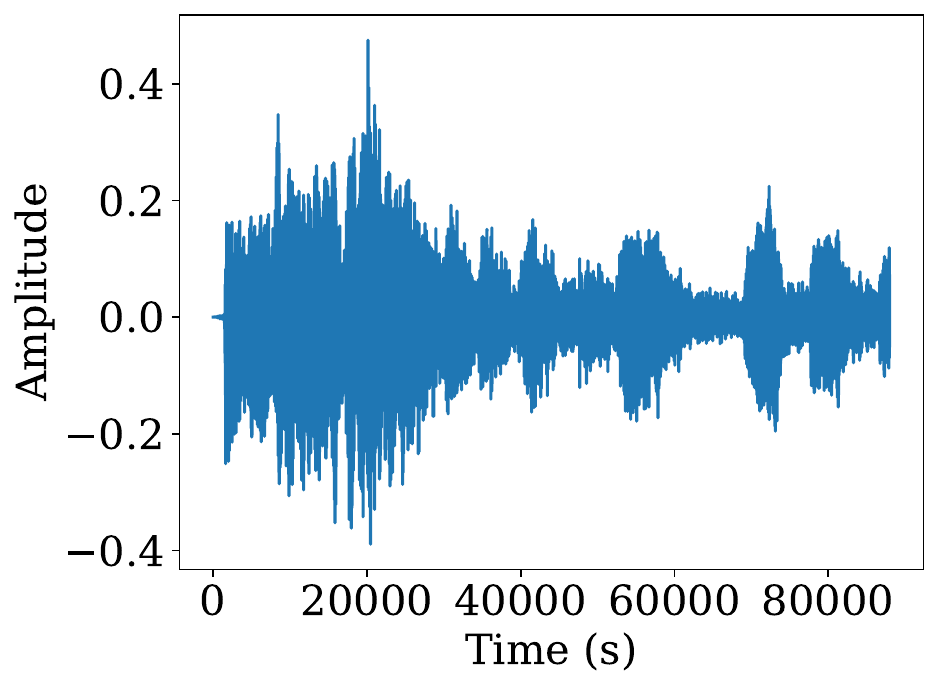}
        \caption{Cover Audio}
        \label{fig:cover_audio}
    \end{subfigure}
    \hfill
    \begin{subfigure}{0.26\textwidth}
        \centering
        \includegraphics[width=\linewidth]{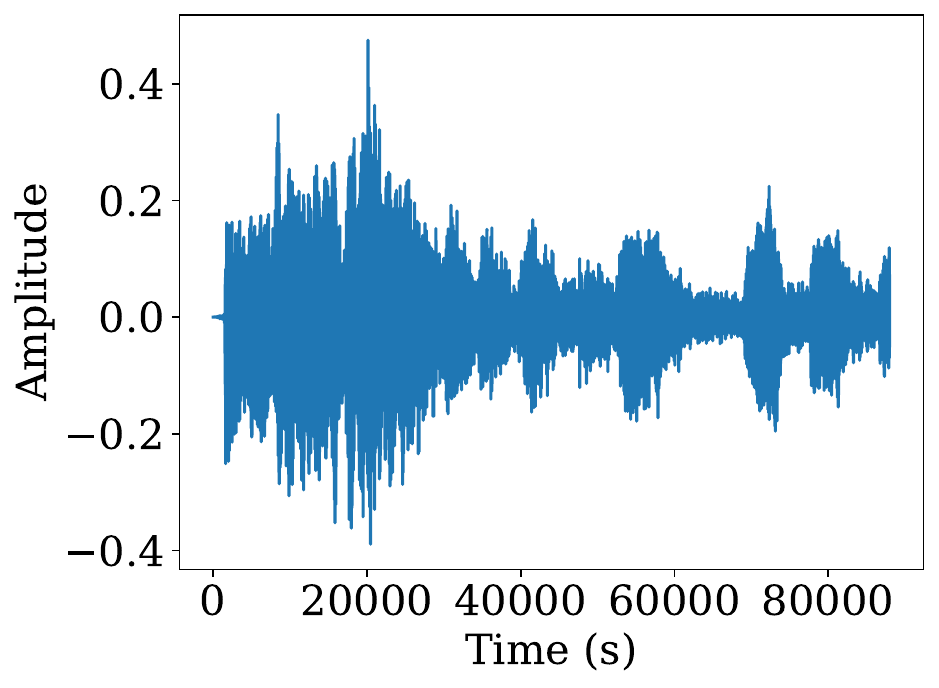}
        \caption{Container Audio}
        \label{fig:container_audio}
    \end{subfigure}
    \hfill
    \begin{subfigure}{0.26\textwidth}
        \centering
        \includegraphics[width=\linewidth]{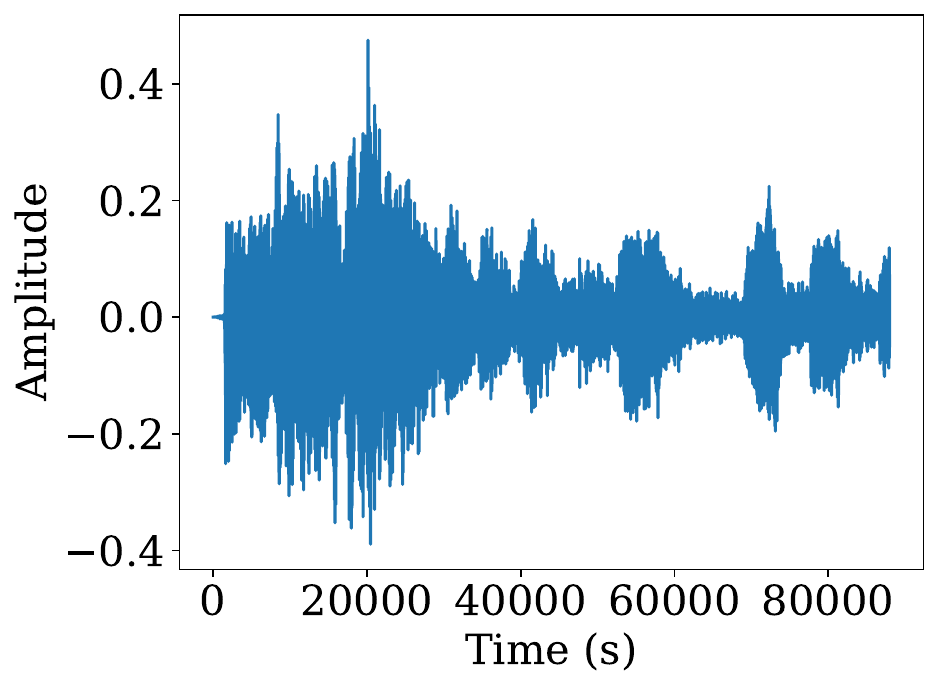}
        \caption{Sanitized Audio}
        \label{fig:sanitized_audio}
    \end{subfigure}
    \vspace{1.0em}
    \caption{A visual comparison between a cover audio, a container audio, and the resulting sanitized audio from DM-SUDS. The container is created by embedding the text secret $\secret$ from using LSB. The container is then sanitized with DM-SUDS to create a sanitized audio.}
    \label{fig:audio_samples}
    \vspace{1.0em}
\end{figure*}

\subsection{Scalability to ImageNet and Robustness (JPEG)}
The results for RQ4 are detailed at the bottom of \tabref{tab:combined_comparison}, under the labels \textit{DDH (ImageNet)} and \textit{PRIS (ImageNet)}. These findings affirm the efficacy of DM-SUDS even when applied to more complex datasets like ImageNet. Notably, the Image Preservation (IP) metrics---low MSE, high PSNR, high SSIM, and high NCC---highlight minimal degradation in image quality, confirming that the visual integrity of the sanitized images is maintained even with more complex images. Additionally, the robust secret elimination (SE) performance is demonstrated by high MSE, low PSNR, low SSIM, and low NCC metric values. These metrics together validate the effectiveness of DM-SUDS in sanitizing more complex datasets and secrets from more robust steganography methods. As $\fullSpec$ is satisfied, both of these methods are successfully sanitized, and \textbf{RQ4 is positive.}

\paragraph{Evaluation} From these experiments, RQ1 and RQ4 are positive and RQ2 and RQ3 are negative. DM-SUDS outperforms previous image-into-image deep learning methods, scales to more complex datasets, and is robust to JPEG-resistant hiding methods (PRIS). Additionally, the power of DM-SUDS lies in the added Gaussian Noise in the forward diffusion process.


%
%


\section{Case Study: Application to Audio Steganography}%
To demonstrate the adaptability of this approach to other domains, we apply DM-SUDS to the sanitization of audio containers. Audio data refers to the representation of sound within a digital system. The process of hiding information in audio is very similar to hiding processes in the image domain. For this case study, we consider LSB and spread spectrum (SS) spatial methods as well as discrete wavelet transform (DWT) transform methods for hiding \textbf{text-into-audio}. 

\textbf{Overview}
We utilize 250 randomly sampled audio files from the UrbanSound8K dataset\footnote{\url{https://urbansounddataset.weebly.com/urbansound8k.html}} to create containers with LSB, SS, and DWT. The UrbanSound8K dataset contains 8732 urban sound clips of less than 4 seconds from 10 different classes, including labels such as `air conditioner' and `street music'. The secret embedded in each audio file is a sentence scraped from Shakespeare's \textit{Twelfth Night}. We use secrets of various content and sizes to stress test the sanitization process. Each of these resulting containers is sanitized using DM-SUDS. Similar to the experiments described in \Secrefs{sec:research_questions}, we collect audio preservation and secret elimination metrics. Audio preservation is evaluated by the MSE difference between the container audio $\container$ and the sanitized audio $\sanitized$. As this case study utilizes text secrets, secret elimination is evaluated using bit error rate (BER) and removal rate (RR) metrics. BER is computed by comparing the bitstream extracted from the sanitized audio with the original embedded bitstream. The removal rate is calculated by \eqrefs{eq:rr} \cite{geng2020real}. If $\text{BER} > 0.2 \implies \text{RR} > 0.4$, the hidden message is successfully destroyed. DM-SUDS for this case study is a publicly available diffusion model trained to generate music\footnote{\url{https://github.com/teticio/audio-diffusion}}. %
\begin{equation} 
\label{eq:rr}
\text{RR} = 1 - || 2\times \text{BER} - 1||
\end{equation}%
%
%
%
\textbf{Results}
From the audio preservation and secret elimination results shown in \tabref{tab:audio_dmsuds_results}, DM-SUDS is also able to sanitize audio containers. The low MSE audio preservation values indicate that the quality of the sanitized audio is preserved during the sanitization process. \figrefs{fig:audio_samples} demonstrates the audio preservation capabilities of DM-SUDS as the sanitized audio (\figrefs{fig:sanitized_audio}) is minimally different from the container (\figrefs{fig:cover_audio}). DM-SUDS is also successful at eliminating secrets. Prior to sanitization, the secret is recovered with a BER of 0.0\%. This means that all secret bits are correctly recovered. After sanitization, however, the revealed secret BER is high, resulting in a high RR. As $\text{BER} > 0.2 \implies \text{RR} > 0.4$, each of the hiding methods (LSB, SS, and DWT) is successfully sanitized with DM-SUDS.%
\begin{table}[]
    \centering
    \caption{Audio sanitization results with DM-SUDS.}
    \vspace{0.75em}
    \begin{tabular}{c|c|c|c|c}
        \toprule
        \multirow{2}{*}{\textbf{$\Hide$}} & \textbf{$\Hide$ Eval} & \textbf{Audio Preservation} & \multicolumn{2}{c}{\textbf{Secret Elimination}} \\
         & BER($\secret, \revealsecret$) &  MSE($\container, \sanitized$) & BER($\secret, \revealsanisecret$) & RR \\
         \midrule
         LSB & 0.0 & 0.0013 & 0.4901 & 0.9273 \\
         SS & 0.0 & 0.0013 & 0.4365 &  0.8647 \\
         DWT & 0.0 & 0.0001 & 0.4851 & 0.9274\\
         \bottomrule
    \end{tabular}
    \label{tab:audio_dmsuds_results}
\end{table}%

\section{Conclusion}
In this work, we introduce a novel blind deep learning steganography sanitization method that utilizes a diffusion model framework to restore images called DM-SUDS. We demonstrate the success of such an approach compared to an existing VAE-based approach (SUDS), Gaussian Noise, and DCT Gaussian Noise and conduct an ablation study to determine the flexibility of the timestep parameter and the ability of DM-SUDS to directly denoise. We also demonstrate the scalability and robustness of DM-SUDS against JPEG-resistant steganography on the ImageNet dataset. Additionally, by incorporating image preservation into a novel sanitization evaluation, we effectively capture the safety and the utility of resulting sanitized images. One of the most impactful features of DM-SUDS lies in its accessibility, as any pretrained diffusion model of a target domain's data distribution can be implemented to protect against steganography. This work, therefore, introduces a highly effective and beneficial use case for a diffusion model while marking a significant advancement in the field of steganography sanitization.  We also demonstrate the successful application of DM-SUDS to sanitizing containers in the audio domain. This is the first work to sanitize secrets hidden in audio containers using deep learning. In the future, we would like to address the speed of sanitization (see \tabref{tab:combined_comparison}), and we would like to incorporate the sanitization or removal of visible watermarks in addition to the invisible information considered in this work.



\begin{ack}
This paper was supported in part by a fellowship award under contract FA9550-21-F-0003 through the National Defense Science and Engineering Graduate (NDSEG) Fellowship Program, sponsored by the Air Force Research Laboratory (AFRL), the Office of Naval Research (ONR), and the Army Research Office (ARO). The material presented in this paper is based upon work supported by the National Science Foundation (NSF) through grant numbers 2220426 and 2220401, the Defense Advanced Research Projects Agency (DARPA) under contract number FA8750-23-C-0518, and the Air Force Office of Scientific Research (AFOSR) under contract number FA9550-22-1-0019 and FA9550-23-1-0135. Any opinions, findings, and conclusions or recommendations expressed in this paper are those of the authors and do not necessarily reflect the views of AFOSR, DARPA, or NSF.
\end{ack}


\bibliography{references}

\begin{thebibliography}{45}
\providecommand{\natexlab}[1]{#1}
\providecommand{\url}[1]{\texttt{#1}}
\expandafter\ifx\csname urlstyle\endcsname\relax
  \providecommand{\doi}[1]{doi: #1}\else
  \providecommand{\doi}{doi: \begingroup \urlstyle{rm}\Url}\fi

\bibitem[Bachrach and Shih(2011)]{bachrach2011image}
M.~Bachrach and F.~Y. Shih.
\newblock Image steganography and steganalysis.
\newblock \emph{Wiley Interdisciplinary Reviews: Computational Statistics}, 3\penalty0 (3):\penalty0 251--259, 2011.

\bibitem[Baluja(2017)]{baluja2017hiding}
S.~Baluja.
\newblock Hiding images in plain sight: Deep steganography.
\newblock \emph{Advances in neural information processing systems}, 30, 2017.
\newblock \url{https://papers.nips.cc/paper/2017/file/838e8afb1ca34354ac209f53d90c3a43-Paper.pdf}.

\bibitem[Baluja(2019)]{baluja2019hiding}
S.~Baluja.
\newblock Hiding images within images.
\newblock \emph{IEEE transactions on pattern analysis and machine intelligence}, 42\penalty0 (7):\penalty0 1685--1697, 2019.
\newblock \url{https://ieeexplore.ieee.org/stamp/stamp.jsp?tp=&arnumber=8654686}.

\bibitem[Cheddad et~al.(2010)Cheddad, Condell, Curran, and Mc~Kevitt]{cheddad2010digital}
A.~Cheddad, J.~Condell, K.~Curran, and P.~Mc~Kevitt.
\newblock Digital image steganography: Survey and analysis of current methods.
\newblock \emph{Signal processing}, 90\penalty0 (3):\penalty0 727--752, 2010.
\newblock \url{https://www.sciencedirect.com/science/article/abs/pii/S0165168409003648}.

\bibitem[Geetha et~al.(2021)Geetha, Subburam, Selvakumar, Kadry, and Damasevicius]{geetha2021steganogram}
S.~Geetha, S.~Subburam, S.~Selvakumar, S.~Kadry, and R.~Damasevicius.
\newblock Steganogram removal using multidirectional diffusion in fourier domain while preserving perceptual image quality.
\newblock \emph{Pattern Recognition Letters}, 147:\penalty0 197--205, 2021.

\bibitem[Geng et~al.(2020)Geng, Zhang, Chen, Fang, and Yu]{geng2020real}
L.~Geng, W.~Zhang, H.~Chen, H.~Fang, and N.~Yu.
\newblock Real-time attacks on robust watermarking tools in the wild by cnn.
\newblock \emph{Journal of Real-Time Image Processing}, 17:\penalty0 631--641, 2020.

\bibitem[Hayes and Danezis(2017)]{hayes2017generating}
J.~Hayes and G.~Danezis.
\newblock Generating steganographic images via adversarial training.
\newblock \emph{Advances in neural information processing systems}, 30, 2017.

\bibitem[Ho et~al.(2020)Ho, Jain, and Abbeel]{ho2020denoising}
J.~Ho, A.~Jain, and P.~Abbeel.
\newblock Denoising diffusion probabilistic models.
\newblock \emph{arXiv preprint arxiv:2006.11239}, 2020.

\bibitem[Holub and Fridrich(2012)]{holub2012designing}
V.~Holub and J.~Fridrich.
\newblock Designing steganographic distortion using directional filters.
\newblock In \emph{2012 IEEE International workshop on information forensics and security (WIFS)}, pages 234--239. IEEE, 2012.

\bibitem[Jing et~al.(2021)Jing, Deng, Xu, Wang, and Guan]{jing2021hinet}
J.~Jing, X.~Deng, M.~Xu, J.~Wang, and Z.~Guan.
\newblock Hinet: deep image hiding by invertible network.
\newblock In \emph{Proceedings of the IEEE/CVF international conference on computer vision}, pages 4733--4742, 2021.

\bibitem[Johnson and Jajodia(1998)]{johnson1998steganalysis}
N.~F. Johnson and S.~Jajodia.
\newblock Steganalysis of images created using current steganography software.
\newblock In \emph{International Workshop on Information Hiding}, pages 273--289. Springer, 1998.

\bibitem[Jung et~al.(2021)Jung, Bae, Choi, and Yoon]{jung2021pixelsteganalysis}
D.~Jung, H.~Bae, H.-S. Choi, and S.~Yoon.
\newblock Pixelsteganalysis: Pixel-wise hidden information removal with low visual degradation.
\newblock \emph{IEEE Transactions on Dependable and Secure Computing}, 2021.

\bibitem[Kishor et~al.(2016)Kishor, Ramaiah, and Jilani]{kishor2016review}
S.~N. Kishor, G.~K. Ramaiah, and S.~Jilani.
\newblock A review on steganography through multimedia.
\newblock In \emph{2016 International Conference on Research Advances in Integrated Navigation Systems (RAINS)}, pages 1--6. IEEE, 2016.
\newblock \url{https://ieeexplore.ieee.org/document/7764373}.

\bibitem[Kurak and McHugh(1992)]{kurak_CSAC}
C.~Kurak and J.~McHugh.
\newblock A cautionary note on image downgrading.
\newblock In \emph{[1992] Proceedings Eighth Annual Computer Security Application Conference}, pages 153--159, 1992.
\newblock \doi{10.1109/CSAC.1992.228224}.
\newblock \url{https://ieeexplore.ieee.org/document/228224}.

\bibitem[Li et~al.(2022)Li, Zeng, Zhang, and Qin]{li2022ensemble}
F.~Li, Y.~Zeng, X.~Zhang, and C.~Qin.
\newblock Ensemble stego selection for enhancing image steganography.
\newblock \emph{IEEE Signal Processing Letters}, 29:\penalty0 702--706, 2022.

\bibitem[Li et~al.(2021)Li, Wang, Ma, Wang, Wang, Gao, and Shi]{li2021concealed}
Q.~Li, X.~Wang, B.~Ma, X.~Wang, C.~Wang, S.~Gao, and Y.~Shi.
\newblock Concealed attack for robust watermarking based on generative model and perceptual loss.
\newblock \emph{IEEE Transactions on Circuits and Systems for Video Technology}, 32\penalty0 (8):\penalty0 5695--5706, 2021.

\bibitem[Li(2023)]{li2023diffwa}
X.~Li.
\newblock Diffwa: Diffusion models for watermark attack.
\newblock In \emph{2023 International Conference on Integrated Intelligence and Communication Systems (ICIICS)}, pages 1--8. IEEE, 2023.

\bibitem[Liu et~al.(2022)Liu, Yang, Jiang, Wu, Peng, Wang, and Wang]{liu2022deep}
Q.~Liu, J.~Yang, H.~Jiang, J.~Wu, T.~Peng, T.~Wang, and G.~Wang.
\newblock When deep learning meets steganography: Protecting inference privacy in the dark.
\newblock In \emph{IEEE INFOCOM 2022-IEEE Conference on Computer Communications}, pages 590--599. IEEE, 2022.

\bibitem[Lu et~al.(2021)Lu, Wang, Zhong, and Rosin]{lu2021large}
S.-P. Lu, R.~Wang, T.~Zhong, and P.~L. Rosin.
\newblock Large-capacity image steganography based on invertible neural networks.
\newblock In \emph{Proceedings of the IEEE/CVF conference on computer vision and pattern recognition}, pages 10816--10825, 2021.

\bibitem[Nichol and Dhariwal(2021)]{nichol2021improved}
A.~Q. Nichol and P.~Dhariwal.
\newblock Improved denoising diffusion probabilistic models.
\newblock In \emph{International Conference on Machine Learning}, pages 8162--8171. PMLR, 2021.

\bibitem[Nie et~al.(2022)Nie, Guo, Huang, Xiao, Vahdat, and Anandkumar]{nie2022diffusion}
W.~Nie, B.~Guo, Y.~Huang, C.~Xiao, A.~Vahdat, and A.~Anandkumar.
\newblock Diffusion models for adversarial purification.
\newblock \emph{arXiv preprint arXiv:2205.07460}, 2022.

\bibitem[Pan et~al.(2022)Pan, Zhang, Zhang, Yan, and Yang]{pan2022house}
X.~Pan, S.~Zhang, M.~Zhang, Y.~Yan, and M.~Yang.
\newblock House of cans: Covert transmission of internal datasets via capacity-aware neuron steganography.
\newblock \emph{Advances in Neural Information Processing Systems}, 35:\penalty0 24838--24850, 2022.

\bibitem[Paul and Mukherjee(2010)]{paul2010image}
G.~Paul and I.~Mukherjee.
\newblock Image sterilization to prevent lsb-based steganographic transmission.
\newblock \emph{arXiv preprint arXiv:1012.5573}, 2010.

\bibitem[Qian et~al.(2018)Qian, Dong, Wang, and Tan]{qian2018feature}
Y.~Qian, J.~Dong, W.~Wang, and T.~Tan.
\newblock Feature learning for steganalysis using convolutional neural networks.
\newblock \emph{Multimedia Tools and Applications}, 77:\penalty0 19633--19657, 2018.

\bibitem[Quan et~al.(2020)Quan, Teng, Chen, and Ji]{quan2020watermarking}
Y.~Quan, H.~Teng, Y.~Chen, and H.~Ji.
\newblock Watermarking deep neural networks in image processing.
\newblock \emph{IEEE transactions on neural networks and learning systems}, 32\penalty0 (5):\penalty0 1852--1865, 2020.

\bibitem[Robinette et~al.(2023)Robinette, Wang, Shehadeh, Moyer, and Johnson]{robinette2023suds}
P.~K. Robinette, H.~D. Wang, N.~Shehadeh, D.~Moyer, and T.~T. Johnson.
\newblock Suds: Sanitizing universal and dependent steganography.
\newblock \emph{European Conference on Artificial Intelligence}, 372:\penalty0 1978--1985, 2023.
\newblock \doi{10.3233/FAIA230489}.

\bibitem[Subhedar and Mankar(2014)]{subhedar2014current}
M.~S. Subhedar and V.~H. Mankar.
\newblock Current status and key issues in image steganography: A survey.
\newblock \emph{Computer science review}, 13:\penalty0 95--113, 2014.
\newblock \url{https://www.sciencedirect.com/science/article/abs/pii/S1574013714000136}.

\bibitem[Tang et~al.(2019)Tang, Li, Tan, Barni, and Huang]{tang2019cnn}
W.~Tang, B.~Li, S.~Tan, M.~Barni, and J.~Huang.
\newblock Cnn-based adversarial embedding for image steganography.
\newblock \emph{IEEE Transactions on Information Forensics and Security}, 14\penalty0 (8):\penalty0 2074--2087, 2019.

\bibitem[Tang et~al.(2020)Tang, Li, Barni, Li, and Huang]{tang2020automatic}
W.~Tang, B.~Li, M.~Barni, J.~Li, and J.~Huang.
\newblock An automatic cost learning framework for image steganography using deep reinforcement learning.
\newblock \emph{IEEE Transactions on Information Forensics and Security}, 16:\penalty0 952--967, 2020.
\newblock \url{https://ieeexplore.ieee.org/document/9205850}.

\bibitem[Trivedi et~al.(2016)Trivedi, Sharma, and Yadav]{trivedi2016analysis}
M.~C. Trivedi, S.~Sharma, and V.~K. Yadav.
\newblock Analysis of several image steganography techniques in spatial domain: A survey.
\newblock In \emph{Proceedings of the Second International Conference on Information and Communication Technology for Competitive Strategies}, pages 1--7, 2016.
\newblock \url{https://dl.acm.org/doi/10.1145/2905055.2905294}.

\bibitem[Vaidya(2023)]{vaidya2023fingerprint}
S.~P. Vaidya.
\newblock Fingerprint-based robust medical image watermarking in hybrid transform.
\newblock \emph{The Visual Computer}, 39\penalty0 (6):\penalty0 2245--2260, 2023.

\bibitem[Volkhonskiy et~al.(2020)Volkhonskiy, Nazarov, and Burnaev]{volkhonskiy2020steganographic}
D.~Volkhonskiy, I.~Nazarov, and E.~Burnaev.
\newblock Steganographic generative adversarial networks.
\newblock In \emph{Twelfth international conference on machine vision (ICMV 2019)}, volume 11433, pages 991--1005. SPIE, 2020.
\newblock \url{https://doi.org/10.1117/12.2559429}.

\bibitem[Wang et~al.(2022)Wang, Wang, and Yang]{wang2022stegnet}
Z.~Wang, G.~Wang, and Y.~Yang.
\newblock Nas-stegnet: Lightweight image steganography networks via neural architecture search.
\newblock In \emph{International Conference on Neural Information Processing}, pages 228--239. Springer, 2022.

\bibitem[Xu et~al.(2016)Xu, Wu, and Shi]{xu2016structural}
G.~Xu, H.-Z. Wu, and Y.-Q. Shi.
\newblock Structural design of convolutional neural networks for steganalysis.
\newblock \emph{IEEE Signal Processing Letters}, 23\penalty0 (5):\penalty0 708--712, 2016.

\bibitem[Xu et~al.(2022)Xu, Mou, Hu, Xie, and Zhang]{xu2022robust}
Y.~Xu, C.~Mou, Y.~Hu, J.~Xie, and J.~Zhang.
\newblock Robust invertible image steganography.
\newblock In \emph{Proceedings of the IEEE/CVF Conference on Computer Vision and Pattern Recognition}, pages 7875--7884, 2022.

\bibitem[Yang et~al.(2023{\natexlab{a}})Yang, Xu, Liu, and Ma]{yang2023pris}
H.~Yang, Y.~Xu, X.~Liu, and X.~Ma.
\newblock Pris: Practical robust invertible network for image steganography.
\newblock \emph{arXiv preprint arXiv:2309.13620}, 2023{\natexlab{a}}.

\bibitem[Yang et~al.(2019)Yang, Ruan, Huang, Kang, and Shi]{yang2019embedding}
J.~Yang, D.~Ruan, J.~Huang, X.~Kang, and Y.-Q. Shi.
\newblock An embedding cost learning framework using gan.
\newblock \emph{IEEE Transactions on Information Forensics and Security}, 15:\penalty0 839--851, 2019.
\newblock \url{https://ieeexplore.ieee.org/abstract/document/8735922}.

\bibitem[Yang et~al.(2023{\natexlab{b}})Yang, Wang, and Zhang]{yang2023general}
Z.~Yang, Z.~Wang, and X.~Zhang.
\newblock A general steganographic framework for neural network models.
\newblock \emph{Information Sciences}, 643:\penalty0 119250, 2023{\natexlab{b}}.

\bibitem[Ye et~al.(2017)Ye, Ni, and Yi]{ye2017deep}
J.~Ye, J.~Ni, and Y.~Yi.
\newblock Deep learning hierarchical representations for image steganalysis.
\newblock \emph{IEEE Transactions on Information Forensics and Security}, 12\penalty0 (11):\penalty0 2545--2557, 2017.

\bibitem[Zhang et~al.(2020{\natexlab{a}})Zhang, Benz, Karjauv, Sun, and Kweon]{zhang2020udh}
C.~Zhang, P.~Benz, A.~Karjauv, G.~Sun, and I.~S. Kweon.
\newblock Udh: Universal deep hiding for steganography, watermarking, and light field messaging.
\newblock \emph{Advances in Neural Information Processing Systems}, 33:\penalty0 10223--10234, 2020{\natexlab{a}}.
\newblock \url{https://proceedings.neurips.cc/paper/2020/file/73d02e4344f71a0b0d51a925246990e7-Paper.pdf}.

\bibitem[Zhang et~al.(2020{\natexlab{b}})Zhang, Chen, Liao, Fang, Zhang, Zhou, Cui, and Yu]{zhang2020model}
J.~Zhang, D.~Chen, J.~Liao, H.~Fang, W.~Zhang, W.~Zhou, H.~Cui, and N.~Yu.
\newblock Model watermarking for image processing networks.
\newblock In \emph{Proceedings of the AAAI conference on artificial intelligence}, volume~34, pages 12805--12812, 2020{\natexlab{b}}.

\bibitem[Zhang et~al.(2018)Zhang, Zhu, Liu, and Liu]{zhang2018efficient}
R.~Zhang, F.~Zhu, J.~Liu, and G.~Liu.
\newblock Efficient feature learning and multi-size image steganalysis based on cnn.
\newblock \emph{arXiv preprint arXiv:1807.11428}, 2018.

\bibitem[Zhang et~al.(2019)Zhang, Dong, and Liu]{zhang2019invisible}
R.~Zhang, S.~Dong, and J.~Liu.
\newblock Invisible steganography via generative adversarial networks.
\newblock \emph{Multimedia tools and applications}, 78:\penalty0 8559--8575, 2019.

\bibitem[Zhou et~al.(2019)Zhou, Feng, Shen, and Zhang]{zhou2019security}
L.~Zhou, G.~Feng, L.~Shen, and X.~Zhang.
\newblock On security enhancement of steganography via generative adversarial image.
\newblock \emph{IEEE Signal Processing Letters}, 27:\penalty0 166--170, 2019.

\bibitem[Zhu et~al.(2018)Zhu, Kaplan, Johnson, and Fei-Fei]{zhu2018hidden}
J.~Zhu, R.~Kaplan, J.~Johnson, and L.~Fei-Fei.
\newblock Hidden: Hiding data with deep networks.
\newblock In \emph{Proceedings of the European conference on computer vision (ECCV)}, pages 657--672, 2018.
\newblock \url{https://link.springer.com/chapter/10.1007/978-3-030-01267-0_40}.

\end{thebibliography}
\end{document}